\documentclass[12pt]{iopart}
\usepackage{graphicx,iopams}

\begin{document}

\title[A two-species atom
interferometer EP test]{How to estimate the differential acceleration in a two-species atom
interferometer to test the equivalence principle }

\author{G. Varoquaux$^1$\footnote{
Present address: Laboratoire de Neuroimagerie Assist\'ee par Ordinateur,
CEA/SAC/DSV/I2BM/NeuroSpin, 91191 Gif-sur-Yvette, France
}, R.A. Nyman$^1$\footnote{
Present address : Centre for Cold Matter, Blackett Laboratory, Imperial College, 
London, SW7 2BW, United Kingdom}, R. Geiger$^1$, P. Cheinet$^1$, A. Landragin$^2$ and P. Bouyer$^1$}
\address{$^1$ Laboratoire Charles Fabry de l'Institut d'Optique,
Campus Polytechnique, RD 128, 91127 Palaiseau, France}
\address{$^2$ LNE-SYRTE, UMR8630, UPMC, Observatoire de Paris,
61 avenue de l'Observatoire, 75014 Paris, France}

\date{\today}
\pacs{37.25.+k, 03.75.{D}g}

\begin{abstract}

We propose a scheme for testing the weak equivalence principle
(Universality of Free Fall) using an atom-interferometric measurement of
the local differential acceleration between two atomic species with a
large mass ratio as test masses. A apparatus in free fall can be used
to track atomic free-fall trajectories over large distances. We show how the differential
acceleration can be extracted from the interferometric signal using Bayesian
statistical estimation, even in the case of a large mass and laser wavelength difference. 
We show that this statistical estimation method does not suffer from
acceleration noise of the platform and does not require repeatable
experimental conditions. We specialize our discussion to a dual
potassium/rubidium interferometer and
extend our protocol with other atomic mixtures. Finally, we discuss the performances
of the UFF test developed for the free-fall (0-g) airplane in the ICE project (\verb"http://www.ice-space.fr") .

\end{abstract}
\submitto{\NJP}
\maketitle

\section{Introduction}

The Einstein equivalence principle
is fundamental to the standard model of particle physics and all metric theories of gravity \cite{lrr-2006-3,Lammerzhal2006}. It can be broken into
three
elementary principles: the local Lorentz invariance; the local position
invariance (also known as universality of red shift); and the
universality of free-fall (UFF), stating that all freely-falling point
particles follow
identical trajectories independent of their internal composition. The Lorentz
invariance and the position invariance are local properties and can be
tested to unmatched precision using atomic clocks
 \cite{Wolf2006,ashby2007,blatt2008} and ultra-stable cavities \cite{muller:050401}. On the contrary, the UFF can only be tested by
tracking trajectories, ideally of freely-falling test masses. Various
extensions to the current theoretical-physics framework predict
violations of the UFF  \cite{Damour1996}. It is thus important to
look experimentally for such violations and push the limits of
experimental tests of the UFF. Moreover, laboratory experiments are an important complement to astrophysical observations in testing fundamental physics, because of the possibility of controlling the environment and repeating the experiments in varying conditions.

In this article, we show that using an atom
interferometer with two different atomic species in free fall can lead to an accurate test of the UFF,
even for different laser wavelength and mass (i.e. different scaling factors for the interferometers). 
We present a protocol that allows us to accurately extract the acceleration difference
and show that this measurement is almost insensitive to strong vibrational
noise or platform movement, which usually limit the atom interferometer accuracy, or 
platform movement. We show how Bayesian statistical methods introduced in \cite{stockton2007} for two identical atom interferometers
can be extended to extract the acceleration
difference between the two atomic species, taking advantage of phase-correlated measurements between both interferometers.
For the sake of clarity,
we focus on the simultaneous use of
rubidium 87 and potassium 39 atoms in a light-pulse gravimeter
 \cite{Kasevich1991}. Comparing the acceleration of these two different
atomic species constitutes a meaningful test of
the UFF, as they combine a large mass ratio (almost a factor of
two), very different nuclear compositions (37 protons and 50
neutrons for $^{87}$Rb and 20/19 for $^{39}$K) and almost equal laser wavelength and thus interferometer scale factors.
We show that, with reasonable experimental parameters, a test
could reach the accuracy of $\eta \sim 5{\cdot} 10^{-11}$ in a one-day parabolic flight campaign in a zero-g airplane, 
and, if extended to long-duration experiments, could compete with the best available apparatus. Our discussion can be generalized to other atomic species
such as the mixture proposed in \cite{dimopoulos2007} or lighter atomic species.

\section{Testing the Universality of Free-Fall}

A figure of merit often used to characterize a test of the UFF is the
E\"otv\"os ratio $\eta$, giving the fractional difference in acceleration
between two test masses in free-fall: $\eta = \Delta a/a$. Alternate
quantum gravitation theories predict deviations from the UFF for $\eta
\lesssim 10^{-13}$  \cite{Damour2007,Sandvik2002,Wetterich2003}. Current
experimental limits on violations of the UFF  ($\eta < 10^{-13}$)
are set by lunar laser-ranging measurements \cite{Williams1996} 
and torsion-balance laboratory experiments  \cite{schlamminger:041101}. Tests of the UFF by monitoring the acceleration
difference between two objects freely falling simultaneously have shown $\eta < 10^{-10}$  \cite{Kuroda1989}.
While all these experiments test the validity of the UFF on macroscopic
objects, in the quest for a quantum gravity theory, it is interesting to
look for deviations on elementary, or microscopic, particles, where quantum mechanics is needed to describe their evolution  \cite{Audretsch1992, Adunas2001}.

The accuracy and sensitivity of local-acceleration measurements using
atom-interferometry  nowadays rival state-of-the-art
conventional accelerometers using macroscopic test masses
 \cite{Kasevich1991, Peters1999, Peters2001}. With such sensors, the quantity measured
directly relates to the acceleration of weakly-interacting particles via experimentally
well-controlled quantities, such as laser wavelengths \cite{Borde1989}. In addition, the evolution of these particles in the gravitational field can be modeled within a covariant quantum field theory  \cite{borde2001b}. Recent results
using atom-interferometric gravimetry to compare the 
acceleration between two isotopes \cite{Peters1999, Fray2004} have demonstrated the
possibility of atom-interferometric tests of the UFF. Ongoing efforts to extend the size of
inertial-sensing atom interferometers by increasing the interrogation
time  \cite{Nyman2006a,dimopoulos2007,STERN:2009:HAL-00369287:1} open the door to high-accuracy atom
accelerometers which will be very sensitive to smaller accelerations,
thus pushing the limits of these tests. These long interrogation times,
i.e. large free-fall heights, can be achieved when using a large
experimental chamber to launch the atoms such as a ten-meter-high
fountain, as suggested in  \cite{dimopoulos2007}.

Compact apparatuses can also be used in reduced-gravity environments,
such as drop towers  \cite{vogel2006,kolnemann2007}, orbital platforms
 \cite{cacciapuoti2007}, or atmospheric parabolic flight
 \cite{STERN:2009:HAL-00369287:1}. However, increasing the interrogation time also
increases the sensitivity of the interferometer to acceleration noise
 \cite{Cheinet2008} which can scale from $\sim 10^{-5}\,{\rm m.s}^2$ in drop
towers to $\sim 10^{-2}\,{\rm m.s}^2$ in the 0-g
Airbus \cite{varoquaux2007a} and on the International Space
Station \cite{Penley2002}. Atoms, isolated in a vacuum chamber, are truly
in free-fall in the Earth's local gravity field, as long as they do not
hit the chamber walls, or experience field gradients (optical or static
magnetic). However, their acceleration is recorded relative to an
ill-defined experimental frame. This can compromise the increase in
sensitivity. Measuring differential phase between similar interferometers
using the same light has been shown to reject common-mode inertial noise up to large
scaling-factors \cite{PhysRevLett.81.971,refId,herrmann2009}, however, in the case of an UFF
measurement, the two interferometers compared do not share the same
sensitivity to inertial effects and the common-mode rejection is not straightforward.

\section{Differential atom interferometer in free-fall}

The acceleration-measurement process on each single species can be
pictured as marking successive positions of freely-falling atoms with a
pair of Raman lasers, pulsed in time. The resulting atomic phase shift
$\phi$ is the difference between the relative phase of the Raman lasers
at the atom's successive classical positions  \cite{Peters1999, borde2001}. When the
Raman lasers are used in a retroreflected configuration, this phase simply relates to the
distance between the atomic cloud and the reference retroreflecting
mirror. In such a three-pulse interferometer, using atoms without initial velocity, the inertial phase shift varies with the acceleration $a$ as:

\begin{equation}
\phi = k\,a\,T^2,
\label{eqn:phi_inertial}
\end{equation}
where $a$ is the acceleration of the atoms relative to the
mirror, $k$ is the effective wavevector of the Raman lasers, and $T$ is
the time between successive pulses.

To test the UFF, we need to extract the difference in
acceleration between the two species. However, in
atom interferometers with internal-state labeling  \cite{Borde1989}, the
experimental signal is the population ratio $n$ between the two output arms of a single
interferometer $n\sim \cos(\phi + \Phi)$,
with $\Phi$ related to the phase noise of the Raman lasers
 \cite{Nyman2006a} and contributions due to vibrations and unwanted
inertial effects on the retroreflecting mirror  \cite{legouet2008}. 
To extract
an absolute value
of the interferometric phase, accumulating data to scan a
fringe is required. This accumulation of several experimental points can thus be
hindered by the acceleration noise of the platform, as the acceleration
measured may vary from one measurement to another. To have access to long integration, we thus want to extract as much information as
possible about the phase difference with a minimum set of independent measurements.

We now focus on the Potassium(K)-Rubidium(Rb)  atom interferometer described in \cite{Nyman2006a}.
The UFF experimental signal that we are interested in
is the E\"otvos ratio, 
$\eta=2(a_{\rm K} -a_{\rm Rb} )/(a_{\rm K} +a_{\rm Rb} )$ with
$a_{\rm K}$ and $a_{\rm Rb}$ the accelerations of potassium and
rubidium. We thus want to extract the acceleration difference $\delta a =
a_{\rm K} -a_{\rm Rb}$. This can be related directly to the difference of
the inertial phase of each interferometer: using Equation
(\ref{eqn:phi_inertial}), $\delta \phi = \phi_{\rm K} - \phi_{\rm Rb}=k_{\rm K}
a_{\rm K}T_{\rm K}^2 - k_{\rm Rb} a_{\rm Rb}T_{\rm Rb}^2$ where
$k_{\rm K} = 4\pi/767\,{\rm nm}^{-1}$ and $k_{\rm Rb} =
4\pi/780\,{\rm nm}^{-1}$ are the effective wave-vectors of the Raman
transitions, and $T_K$ and $T_{\rm Rb}$ are the interrogation times for
the K and Rb interferometers respectively. In order to directly read out $\delta a$, we need to adjust the
respective interferometer interrogation times so that they have the same
scale factor: $k_{\rm Rb}T_{\rm Rb}^2=k_{\rm K}T_{\rm K}^2={\cal S}$,
i.e. $T_{\rm Rb}/T_{\rm K} \sim 1.008$. In this case, we simply have 
\begin{equation}
\delta
\phi = {\cal S} \delta a.
\label{diff_meas}
\end{equation}

As in \cite{stockton2007} we make a statistical
description of the measurement process in a two-species interferometer.
Two quantities are measured:
\begin{equation}
\cases{
n_{\rm K}\,\,=A + B \cos ( \phi_{\rm K} + \Phi_{\rm K} )\\
n_{\rm Rb}=C + D \cos ( \phi_{\rm Rb} + \Phi_{\rm Rb} )\\
}
\label{eqn:model1}
\end{equation}
where the capital letters represent fluctuating quantities: $A$ and
$C$ are the offsets of the population measurement, $B$ and $D$ are the
fringe visibilities, and $\Phi_{\rm K}$ and $\Phi_{\rm Rb}$ are related to the
phase and acceleration noises on the two interferometers. 
In the following, we
neglect the laser-induced phase noise, due e.g. to finite laser linewidth
 \cite{legouet2007} or microwave source jitter  \cite{Nyman2006a}, since
it can be reduced with appropriate phase-locking techniques  \cite{legouet2008}. 

The
interferometric phase noise, due to vibrations or other uncontrolled
inertial effects, appears as shifts of the local Raman phase for each
interferometer. Following \cite{cheinetphd} and  assuming a white acceleration noise
of  power spectral density (PSD) $S_\alpha^0$, the standard deviation of $\Phi_{\rm K,Rb}$ can be written 
\begin{equation}
\sigma_{\Phi_i}=k_i {T_i^{3/2}}\sqrt{\frac{2 S_\alpha^0}{3}}, i={\rm K,Rb}.
\end{equation}
This phase noise can be expressed as a random spatial displacement $X(t)$ of deviation $\sigma_X=\sigma_{\Phi}/k$
of the retroreflecting mirror.
The effect of other common-mode, spatial, Raman-phase
fluctuations such as optical aberrations  \cite{fils2005} can also be included
in this fluctuating variable $X$. 

When calculating the differential response of the two-species atom interferometer, 
the relative phase noise between the two interferometers can be written
$\Delta \Phi = \Phi_{\rm K}-\Phi_{\rm Rb} = \,k_{\rm K} \, \tilde{X}$
where the standard deviations of $\tilde{X}$  and $X$ are simply related by a vernier-scale relation 
\begin{equation}
\label{eq:vernier}
\sigma_{\tilde X}= \sigma_X \frac{\delta k}{2k}
\end{equation}
with $\sigma_X$ calculated for $T=T_{\rm K}$ and $\delta k/2k=(k_{{\rm K}}-k_{{\rm Rb}})/2k_{{\rm K}}\approx 8.5{\cdot}10^{-3}$. The derivation of this scale relation and its physical explanation are detailed in \ref{appendixA}.
We can now rewrite Eq.\,(\ref{eqn:model1}) as a simplified measurement model, highlighting
differential effects:
\begin{equation}
\cases{
n_{\rm K} \,\,= A + B \cos \bigl(
    \tilde{\Phi}_{\rm Rb}+ \delta\phi
    + k_{\rm K} \, \tilde{X}
    \bigr)
\\
n_{\rm Rb} = C + D \cos \tilde{\Phi}_{\rm Rb} }
\label{eqn:model2}
\end{equation}
The phase $\tilde{\Phi}_{\rm Rb}={\cal S}
a_{\rm Rb}+\Phi_{\rm Rb}$ of the rubidium interferometer is taken as the
reference. The phase offset in the potassium interferometer
includes the UFF signature, $\delta\phi$ and the
effect of vibrations of the retroreflecting mirror, $k_{\rm K} \,
\tilde{X}$.

\section{Extracting the differential phase using Bayesian analysis}

\begin{figure}
\includegraphics[width=0.9\linewidth]{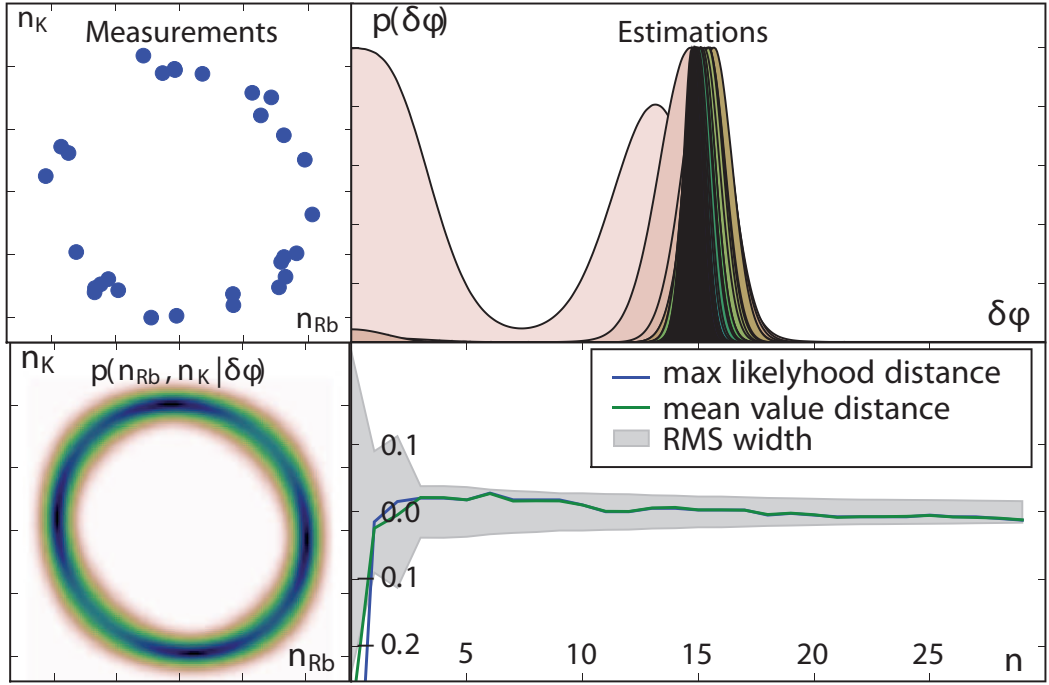}

\caption{Simulation of the model of Eq.\,(\ref{eqn:model2}). The random variable $\tilde{\Phi}_{\rm Rb}$ is 
generated from a uniform distribution taken in the interval $0, 2\pi$. We generate $X$ 
from a normal distribution centered around 0 with $\sigma_X = 1.2$
$\mu$m. $A,B,C$ and $D$ are generated from a normal distribution with
$\sigma = 0.05$. The mean values are 0 for $A$ and $C$ and $1$ for $B$ and
$D$. For each simulation are displayed (going clockwise, starting bottom
left) : the measurement probability in the $(n_{\rm Rb}, n_{\rm K})$
plane for the value of the hidden parameter $\delta\phi$ choosen in the 
simulation, the different measurements actually drawn, the successive
probability-distribution estimations for $\delta\phi$, and the distance 
between estimated values of $\delta\phi$ and the actual value.
}

\label{convergence_example}

\end{figure}

To extract the differential phase $\delta \phi$ from statistically independent data
acquired during different measurement sequences, we use
recursive Bayesian estimation  \cite{Wasserman2004,Zawisky1998}. Although
several solutions for signal
processing in atom-interferometric inertial sensors have been studied
 \cite{stockton2007,foster2002}, Bayesian estimators make best use of
a noise model for estimating information from experimental data
 \cite{stockton2007} and they have been successfully used in quantum optics  \cite{Zawisky1998,Guerlin2007} 
 or in optical interferometers  \cite{Dupuis2005}.
Moreover, Kalman filtering, a restricted version of Bayesian estimation,
plays a critical role for proper use of fiber-optic gyroscope data
 \cite{Gaiffe2002}.
In the Bayesian framework, the parameters to be estimated are considered as
random variables, whose probability distribution is deduced from the
measurements by inverting the measurement model.
Using Bayes' theorem, we calculate the probability distribution function $p(\delta \phi \, | \,
{\scriptstyle (n_{\rm K}, n_{\rm Rb})}_i )$ for the
parameter $\delta \phi$ given the results of a coupled measurement on both
interferometers $(n_{\rm K}, n_{\rm Rb})_i$,
for each measurement $i$. The probability distribution $p(\delta \phi \,
    | \, {\scriptstyle (n_{\rm K}, n_{\rm Rb})}_1, \,
         {\scriptstyle (n_{\rm K}, n_{\rm Rb})}_2, \dots )$ for $\delta \phi$,
given all measurements, is the product of all these
conditional probabilities:
\begin{equation}
p(\delta \phi \, | \, {\scriptstyle (n_{\rm K}, n_{\rm Rb})}_1, \,
         {\scriptstyle (n_{\rm K}, n_{\rm Rb})}_2, \dots ) = 
p(\delta \phi \, | \, {\scriptstyle (n_{\rm K}, n_{\rm Rb})}_1 ) \;
p(\delta \phi \, | \, {\scriptstyle (n_{\rm K}, n_{\rm Rb})}_2 ) \dots
\label{eqn:bayesian_estimator}
\end{equation}
The non linearities of the measurement model introduced by the
trigonometric functions in Eq.\,(\ref{eqn:model2}), with different periods $k_{\rm Rb}$ and
$k_{\rm K}$, make the analytical calculation of the conditional
probability required for the Bayesian estimation tedious and the
resulting expression is computationally costly to evaluate. We calculate
this probability distribution (i.e. the ``posterior'')  by using Monte Carlo sampling of
the state space with the noise model  (Eq.\,(\ref{eqn:model2})).
We use the probability law to estimate the reversed conditional
probability (i.e. the ``prior'') $p(n_{\rm K}, n_{\rm Rb} | \delta \phi)$. A
kernel density estimator  \cite{Wasserman2004} can be used to reduce the
number of sampling points required, though this may not increase
overall numerical efficiency. The posterior, used for
Eq.\,(\ref{eqn:bayesian_estimator}), is
obtained from the prior using Bayes' theorem \cite{Bayes}.

\begin{figure}
\includegraphics[width=0.9\linewidth]{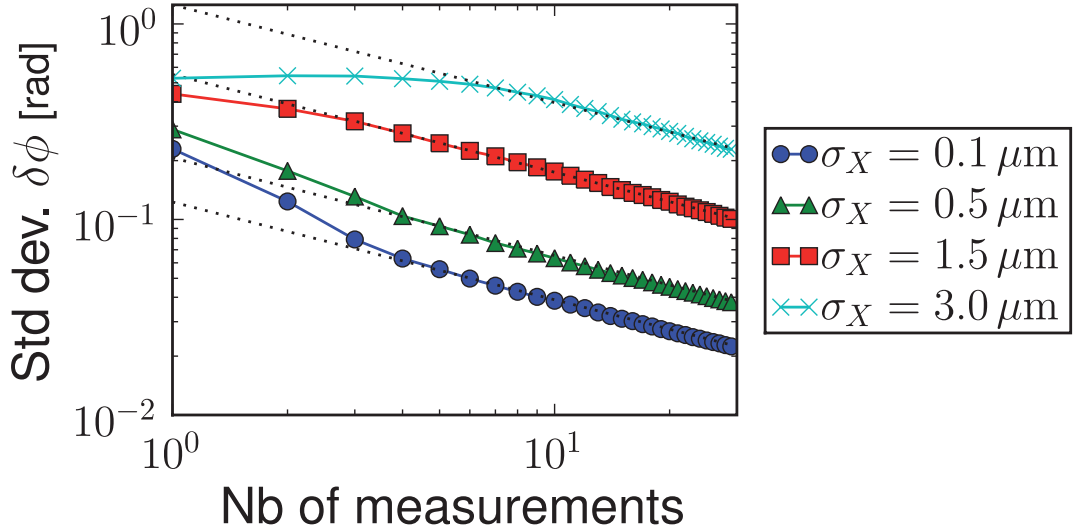}

\caption{Standard deviation of the Bayesian estimation
of the differential phase $\delta \phi$ between K and Rb interferometers.
The standard deviations of the estimation of the differential phase decrease at different rates  with
repeated measurements for different
mirror-displacement amplitudes (dotted lines). The convergence of the points towards the dotted lines shows how correlated noise is handled by the algorithm.}
\label{convergence_speed}

\end{figure}

We have run Monte Carlo simulations of the estimation process for
different values of displacement noise amplitude $\sigma_X$ with the measurement
model described in Eq.\,(\ref{eqn:model2}), as pictured in figure \ref{convergence_example}. We use Gaussian noise
of standard deviation $\sigma = 0.05$ for the various
parameters. The
common-mode phase $\tilde{\Phi}_{\rm Rb}$ is taken to be uniformly
random in the interval $[0, 2\pi]$. 
As seen in figure \ref{convergence_speed}, the estimator converges to a precise
value of the differential phase with fewer than 10 uncorrelated measurements
for fluctuations of $X$ up to a few micrometers, which corresponds to a drift
of the one-species interferometric signal by several complete fringes.
This surprising result comes from the vernier scale relation in Eq.\,(\ref{eq:vernier}) that reduces the differential measurement sensitivity
to Raman-phase fluctuations.

One should
note that if $\delta \phi \sim 0$, the measurements performed do not
contain enough information for good estimation (we have
checked that the Fisher information matrix  \cite{Wasserman2004} is zero
for $\delta \phi = 0$). This can be easily understood 
since the data  $n_{\rm K}$ and $n_{\rm Rb}$ are then distributed along a single line and we lose
the sensitivity to phase/vibration noise \cite{foster2002}.
It is thus necessary to introduce a controlled
phase jump of $\pi/2$ in one of the two interferometers, e.g. on the
phase of the Raman lasers. This choice of interferometric phase shift 
corresponds to working on the side of a fringe in a standard interferometer:
the two parameters are maximally independent, distributed on a 
large-ellipticity curve rather than a flat line (see figure 
\ref{convergence_example}), giving optimal measurement
sensitivity.

We have performed other simulations for different choices of noise
distribution for $X$, which can be related to acceleration-noise power spectral density using the
interferometer sensitivity function as described in \ref{appendixA}. We find that the convergence rate is not dependent
on the nature of the bell-shaped noise distribution, nor on its behavior
in its wings, but only on the RMS amplitude of the fluctuations. In
addition, we have performed simulations for different dual-interferometer
configurations in which the distance between the effective Raman
transition wave vectors is reduced, thus reducing the Vernier-scale
effect. Specifically we have investigated using the potassium D1
transition (at $770\,{\rm nm}$) with rubidium, or the two isotopes of
rubidium, with transitions separated only by $3\,{\rm GHz} \sim
0.03\,{\rm nm}$ (see tab. \ref{tab:convergence_speed}). We find that for
small  variations of $X$, ($\sigma_{\rm X} <0.5\,\mu{\rm m}$ typ.), the
phase-estimation is not limited by the Vernier scale effect as the
convergence speed is similar for all three intereferometer
configurations.

\begin{table}
\begin{tabular}{@{\extracolsep{1.8ex}}rccc}
& K--D2{\large\bfseries /}Rb
& K--D1{\large\bfseries /}Rb
& $^{85}$Rb{\large\bfseries /}$^{87}$Rb
\\
\hline \hline
$\sigma_X \sim 0.1\,\mu{\rm m}$
& $.022$
& $.022$
& $.021$
\\
$\sigma_X \sim 0.5\,\mu{\rm m}$
& $.037$
& $.033$
& $.021$
\\
$\sigma_X \sim 1.5\,\mu{\rm m}$
& $.101$
& $.075$
& $.021$
\\
$\sigma_X \sim 3.0\,\mu{\rm m}$
& $.229$
& $.166$
& $.022$
\end{tabular}

\caption{
Standard deviation on the phase estimate, in radians, after 30
measurements, for different mirror-displacement amplitudes
$\sigma_{\rm X}$, and for different pairs of Raman transitions. The
species and transition lines considered here are:
the potassium D2 line at $767\,{\rm nm}$, the potassium D1 line at
$770\,{\rm nm}$, and the rubidium 85 and 87 D2 lines
at $780\,{\rm nm}$, with a distance of $3\,{\rm GHz} \sim
0.03\,{\rm nm}$ between the two isotopes.
\label{tab:convergence_speed}}

\end{table}

\section{Higher-order inertial effects}

We now consider the contribution of additional inertial effects on the
differential measurement strategy. Equation (\ref{eqn:phi_inertial}) gives only dependence of the phase shift to an acceleration, 
and does not include rotation, gravity gradients, and higher-order effects. 
The effect of rotation can, in principle, be rejected by a
feed-forward on the phase of the lasers \cite{Yver-Leduc2003} and a accurate control of the atom cloud velocity after release. 
We are then left with the major second-order
contribution to the phase shifts, the effect of the gravity gradients. Since the atomic trajectories in the two interferometers explore
slightly different altitudes, gravity gradients will add a contribution to the inertial phase shift
 \cite{wolf1999,Borde2002}:
\begin{equation}
\phi_\gamma \sim k \, \gamma \, T^2 \biggl(
\frac{7}{12} T^2 a - \bigl(v - \frac{v_r}{2} \bigr) T \biggr),
\label{meas_grad}
\end{equation}
where $v$ is the relative velocity between the center of mass of the atom cloud and the reference mirror. $v_r=\frac{\hbar k}{m}$ is the recoil velocity and $\gamma$ the gravity gradient.
If the scale factors $\cal S$ of both interferometers are kept constant, the
numerical value of the total differential phase shift (Eq.\,\ref{diff_meas})
then becomes $\delta\phi_{\rm tot}=\delta\phi+ \delta\phi_\gamma$
with $ \delta\phi_\gamma$ the residual gradient-induced phase shift. In 
general, the contribution to $ \delta\phi_\gamma$ of first term in Eq.\,(\ref{meas_grad}) is negligible and $v$ can be chosen to compensate for the recoil velocity $\hbar k / m$.

\section{Conclusion}

We finally turn to the specific case of the experiment in the 0-g
airplane described in \cite{STERN:2009:HAL-00369287:1}.
We can use the results from our simulation to give an order of magnitude
of the precision achievable by a campaign of measurements in Zero-G flights. For
a residual displacement from free fall $X$ such as $\sigma_X \sim 1\,\mu{\rm m}$ (achieved with proper decoupling from the low frequency vibrations), 
 after 30 data points, the standard
deviation on $\delta \phi$ is $\sim 3{\cdot}10^{-2}\,{\rm rad}$. Thus, for
an interrogation time of
$T=2\,{\rm s}$, the differential-acceleration resolution is $\sim
2.5{\cdot}10^{-10}\,{\rm m}{\cdot}{\rm s}^{-2}$, and the $\eta$
parameter characterizing a test of the UFF can be measured to a precision
of $\eta \sim 5{\cdot}10^{-11}$. 
For the experimental conditions encountered in a 0-g plane, an initial
velocity $v \leq 1\,{\rm cm}{\cdot}{\rm s}^{-1}$ and
$T=2\,{\rm s}$, the effect of gravity gradients is 
negligible. 

In conclusion, we have shown that Bayesian
estimation can be efficient to perform a differential
measurement between two inertial sensors using atoms of different mass and interrogation wavelength.
Even for large vibrational noise, and large interrogation times, the Bayesian estimator
converges rapidly. The measurement of the differential phase shift , i.e. the acceleration difference can thus be measured to a high
precision. This opens new perspectives for the development of high precision test of fundamental physics
such as tests of the equivalence principle.
For example, we predict a precision 
of $\eta \sim5{\cdot}10^{-11}$ when using only 30 experimental data points with a free-fall time of $4\,{\rm s}$
in the Zero-G Airbus, such as for the ICE experiment  \cite{Nyman2006a,STERN:2009:HAL-00369287:1}. In the future, free-fall and integration times may be increased by
deploying atom-interferometric inertial sensors on dedicated orbital platforms for
next-generation tests of the UFF, at the price of an increased sensitivity to vibrational noise. The use of fast-convergence
estimators will help rejecting this acceleration noise and thus relax the requirement on
drag-free vibration isolation performance. A rough estimate indicates that for 20 seconds of
interrogation time and an integration over 1 year  \cite{Ertmer2008}, a target
accuracy of $\eta \sim 8{\cdot}10^{-15}$, close to that of the project $\mu$SCOPE
 \cite{touboul2001}, is reachable with no specific drag-free platform.

\ack

The authors would like to thank Jonas Kahn and Peter Wolf for enlightening discussions. The I.C.E. collaboration is funded by the
Centre National d'Etudes Spatiales (CNES), IFRAF and RTRA ``Triangle de la Physique''. Further support
comes from the European Union STREP consortium FINAQS and the European
Space Agency MAP program A0-2004-064/082.

\appendix
\section{Simple derivation of the single-atom interferometer spatial-displacement noise}

It can be shown \cite{Peters2001,storey1994} that the major contribution to the interferometric phase shift is due to the interaction 
with the Raman beams. Whenever the state of the atom changes during such an interaction, it acquires an 
additional phase $\phi_i (\xi_i,t_i)= k\,\xi_i - \omega\,t_i$. The sign of the phase depends on the initial state of the atom. The 
position of the atom $\xi_i$ with respect to the retroreflecting mirror, taken at the time $t_i$ of the pulses $i=\{1,2,3\}$, can be written 
$\xi_i=\xi(t_i)=x(t_i)+{\cal X}(t_i)$ where $x(t)$ is the absolute position of the atom and ${\cal X}(t_i)$ is a random mirror position related to the vibration noise.
Taking $\xi_2=t_2=0$, $t_3=-t_1=T$
and tracing all the 
state changes leads to a phase difference
\begin{equation}
\phi = k\,[x(-T)+x(T)]+k\,[{\cal X}(-T)+{\cal X}(T)]
\end{equation}

Without gravitational field the trajectories are straightlines and the inherent symmetry of the situation 
leads to $\phi = k\,[{\cal X}(-T)+{\cal X}(T)]$. Here, the interferometric phase shift is only related to the vibrational noise and can be written
$\phi=\Phi = k\,X(T)$ where $X(T)$ is a random variable representing the amplitude of the vibrational noise phase shift $\Phi$ for an interrogation time $T$, as calculated in \ref{appendixA}.

The introduction of a gravitational field breaks the symmetry. The atom now 
falls three times as far during transit in the second half of the interferometer as in the first half and we 
find an additional contribution to the phase shift $\phi=k\,a\,T^2$ proportional to the gravitational acceleration,
so that the interferometric signal becomes :
\begin{equation}
n\,=A + B \cos ( \phi+ \Phi )
\end{equation}

\section{Interferometric phase noise of the two-species accelerometer}
\label{appendixA}

In this section, we derive the vernier scale relation in Eq.\,(\ref{eq:vernier}) between the interferometric phase noise standard deviations of the two-species (K-Rb) and one-species (Rb) accelerometers. Following \cite{Cheinet2008,cheinetphd}, the interferometric phase noise can be written

\begin{equation}
\label{eq:B1}
\Phi=\int_{-\infty}^{+\infty} h(t) \frac{d(k\,{\cal X}(t))}{dt} dt,
\end{equation}
where $h(t)$ is the sensitivity function of the interferometer (see reference \cite{Cheinet2008} for its definition) and ${\cal X}(t)$ represents the retroreflecting mirror position at time $t$ so that $d^2{\cal X}(t)/dt^2=\alpha(t)$ with
$\alpha(t)$  the acceleration noise of the experimental platform. Integrating by parts equation (\ref{eq:B1}) leads to

\begin{equation}
\label{eq:parts}
\Phi=\Big[f(t) \frac{d(k\,{\cal X}(t))}{dt} \Big]_{-\infty}^{+\infty} -  \int_{-\infty}^{+\infty} k\, f(t)  \alpha(t) dt, 
\end{equation}
with $k\, f(t)=\int_{0}^{t} k\, h(u) du$. Neglecting the duration of the Raman light pulses ($\tau\sim20\mu s$ typically) with respect to the interrogation time ($T\sim2s$), the function $h(t)$ is an odd and piecewise constant function  and the function $f(t)$ is an even and piecewise linear function; moreover, $h(t)$ and $f(t)$ are  equal to zero out of the window $[-T,T]$. Thus, the first term in the above equation vanishes and the variance of the interferometric phase noise can be written:

\begin{equation}
\label{eq:phi2}
\sigma_{\Phi}^2=\langle\Phi^2 \rangle= \int\hspace{-8pt}\int_{-\infty}^{+\infty}dt_1 dt_2\, k f(t_1) k f(t_2) \, \langle \alpha(t_1) \alpha(t_2) \rangle.
\end{equation}

\begin{figure}[h]
\includegraphics[width=1.0\linewidth]{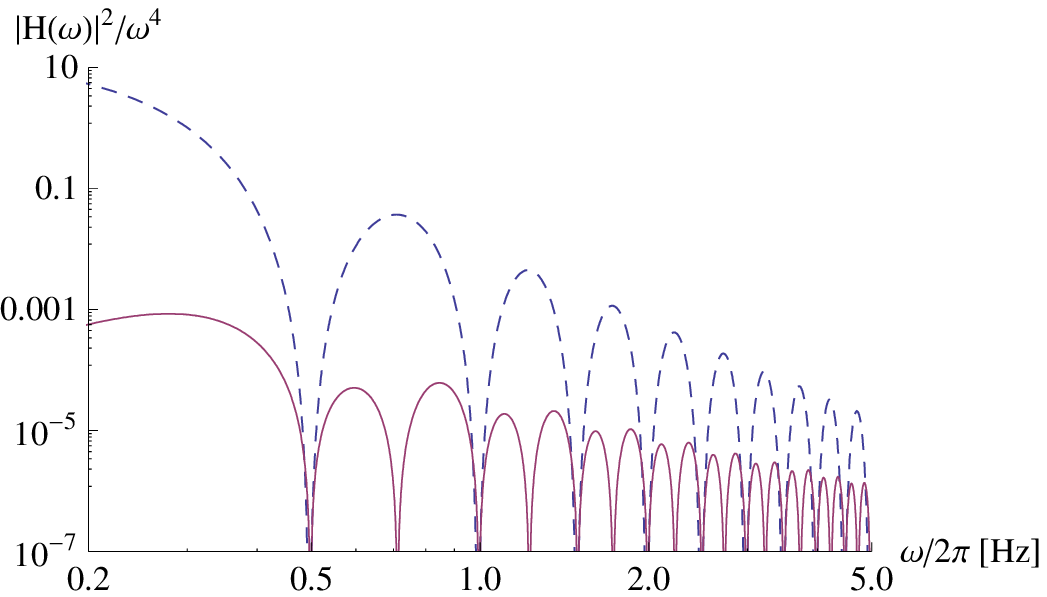}
\caption{Comparison of the transfer functions of the two-species (K-Rb) and one-species (Rb, dashed line) accelerometers in the low frequency range for $T=2$ s.}
\label{HRb_Hdiff_low_freq}
\end{figure}

Assuming a white acceleration noise, i.e. $\langle \alpha(t_1) \alpha(t_2) \rangle= S_\alpha^0 \, \delta(t_1-t_2)$ results in a major simplification of equation (\ref{eq:phi2}), and a straightforward single-integration of $f(t)^2$ finally leads to:
\begin{equation}
\sigma_{\Phi}^2=(\sigma_{X}/k)^2=2 k^2 T^3 S_\alpha^0/3
\label{apeneq1}
\end{equation}

In the case of a differential acceleration measurement, the two interferometers have different sensitivities $h_{{\rm K}}(t)$ and $h_{{\rm Rb}}(t)$ (because $T_{{\rm K}}\ne{T_{\rm Rb}}$). Additionally, the impact of the acceleration noise is slightly different on the K-interferometer ($\propto k_{{\rm K}}{\cal X}(t)$) and on the Rb-interferometer ($\propto k_{{\rm Rb}}{\cal X}(t)$). The relative interferometric phase noise $\Delta \Phi = \Phi_{\rm K}-\Phi_{\rm Rb}$ can be easily calculated by replacing $k f(t)$ by $k_{\rm K} f_{\rm K}(t) - k_{\rm Rb} f_{\rm Rb}(t)$ in Eq.\,(\ref{eq:phi2}) . We find to leading order in $\delta k/k\approx 0.017$:
\begin{equation}
\label{eq:delta_phi_diff}
\sigma_{\Delta\Phi}^2=\frac{2 k^2 T^3 S_\alpha^0}{3} \Big(\frac{\delta k}{2k}\Big)^2 + \mathcal{O}\Big(\frac{\delta k}{k}\Big)^3.
\end{equation}
Combining Eq.\,(\ref{eq:delta_phi_diff}) and Eq.\,(\ref{apeneq1}), we can estimate the vibration noise rejection ratio between the one-species and the two-species interferometer:
\begin{equation}
\label{eq:vernier_tempo}
\frac{\sigma_{\tilde{X}}}{\sigma_X}=\frac{k_{{\rm K}}\,\sigma_{\Delta\Phi}}{k_{{\rm K}}\,\sigma_{\Phi}}\approx\frac{\delta k}{2k}\approx 0.00852.
\end{equation}

It is also interesting to visualize this noise rejection of the two-species interferometer in the frequency domain since one often has access to the acceleration noise spectrum of the experimental platform. In this formalism, the interferometer sensitivity functions $H_{{\rm K}}(\omega)$ and $H_{{\rm Rb}}(\omega)$ result from the Fourier transforms of $h_{{\rm K}}(t)$ and $h_{{\rm Rb}}(t)$. The variance of the relative interferometric phase noise can be written: 
\begin{equation}
\label{eq:vernier_numeric}
\sigma_{\Delta\Phi}^2=\int_0^{+\infty} |H(\omega)|^2 \frac{S_\alpha(\omega)}{\omega^4} \frac{d\omega}{2\pi},
\end{equation}
with $S_\alpha(\omega)$ the acceleration noise PSD and $H(\omega)=H_{{\rm K}}(\omega) - H_{{\rm Rb}}(\omega)$ the transfer function of the differential interferometer. In figure \ref{HRb_Hdiff_low_freq}, we have plotted $|H(\omega)|^2/\omega^4$ and the transfer function $|H_{\rm Rb}(\omega)|^2/\omega^4$ of a one-species (Rb) interferometer, for perfect $\pi/2$ pulses, an effective Rabi frequency $\Omega=2\pi\times50{\rm kHz}$ of the Raman transitions and $T=2$ s. This figure shows that the relative phase noise $\Delta\Phi$ is considerably reduced with respect to the phase noise of a single interferometer in the low frequency domain. This feature is all the more interesting that the amplitude of the acceleration noise of the platform (such as the Zero-G Airbus) is usually the highest at low frequencies.

\section{Noise rejection ratio in the aircraft A300-0G experimental platform}

In this appendix we evaluate the noise rejection of the two-specices accelerometer in the experimental platform where the I.C.E. experiment is performed, namely the A300-0G aircraft carrying out parabolic flights. We take into account the measured acceleration noise spectrum $S_\alpha(\omega)$ in the plane during the quiet part of the parabola
, in the direction of the Raman beams propagation. With the notations introduced in Appendix B, the noise rejection ratio is now given by

\begin{equation}
\label{rejection_ratio}
\frac{\sigma_{\tilde{X}}}{\sigma_X}=\frac{\sigma_{\Delta\Phi}}{\sigma_{\Phi}}= \Big[ \frac{\int_0^{+\infty} d\omega \, |H(\omega)|^2 S_\alpha(\omega)/\omega^4}{\int_0^{+\infty} d\omega \, |H_{{\rm K}}(\omega)|^2 S_\alpha(\omega)/\omega^4} \Big]^{1/2},
\end{equation}
with $H(\omega)$ the transfer function of the differential accelerometer ($H_{{\rm K}}(\omega)$ for the K-accelerometer).

\begin{figure}[h]
\includegraphics[width=.5\linewidth]{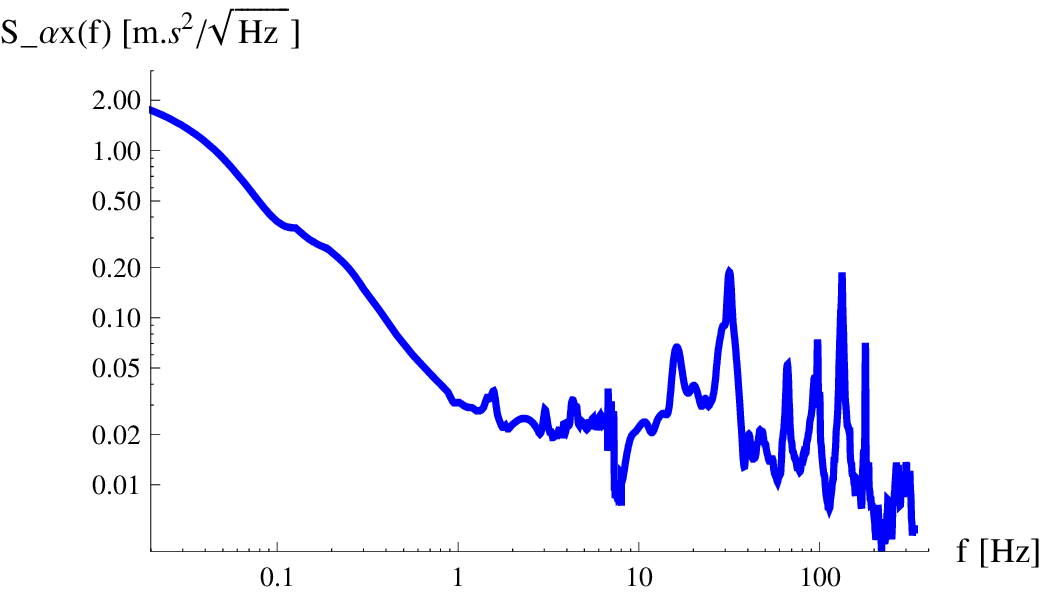}\hskip.1cm\includegraphics[width=.5\linewidth]{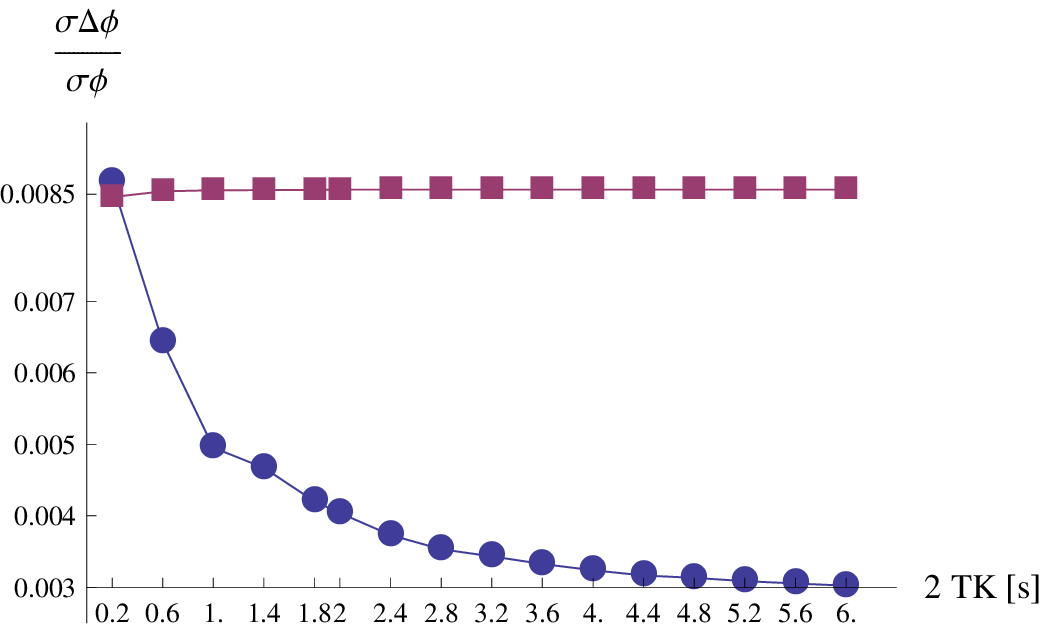}
\caption{Left : measured acceleration noise spectrum of the A300-0G aircraft in the direction of the Raman beams. Right : noise rejection ratio between the one-species and the differential accelerometers for various interrogation times $T_{{\rm K}}$ (various scale factors). The squares correspond to a white acceleration noise whereas the dots account for the real noise spectrum in the plane.}
\label{rejection_TK}
\end{figure}

In figure \ref{rejection_TK}, we have plotted the evolution of the rejection ratio with the total interrogation time of the interferometer. We obtained these data by numerical integrations in equation (\ref{rejection_ratio}), performed on one hand for the real acceleration noise in the plane, and on the other hand for a white noise ($S_\alpha(\omega)=S_\alpha^0$). In the case of a white acceleration noise, we find that the noise rejection ratio is independant of the interrogation time and is equal to the vernier factor $\delta k/2k\approx 0.0085$, as it was derived in \ref{appendixA}.
This rejection ratio can be seen as the weighted average rejection on a limited bandwidth around $1/T$ which contributes the most to the interferometric phase noise. When $T$ is increased, this bandwidth shifts to lower frequencies together with the rejection efficiency. On a white noise, this leads to a constant rejection ratio. On a structured noise such as the real noise spectrum measured in the plane (figure  \ref{rejection_TK}), this is no longer true. For $T_K=100$ ms, the relevant bandwidth is around $10$ Hz where the spectrum is relatively flat and we find a rejection similar to the white spectrum case. For larger values of $T_K$, the noise spectrum gives more weight to the low frequencies for which rejection is more efficient. The rejection ratio is then improved up to a factor 3 as one can see in figure \ref{rejection_TK}. In other words, both accelerometers (one species and two-species) operate where the noise is stronger for longer $T$, but the two-species accelerometer phase noise will increase less than the one-species one.

\end{document}